\documentclass[iop]{emulateapj}

\usepackage{lineno}

\usepackage{xcolor}
\usepackage{amsmath}
\usepackage{graphicx}
\usepackage{amssymb}
\usepackage{psfrag}
\usepackage{dcolumn}
\usepackage{bm}

\usepackage{natbib,times}
\newcommand{\beq}{\begin{equation}}
\newcommand{\eeq}{\end{equation}}

\newcommand{\rmd}{\mathrm{d}}
\newcommand{\brac}[1]{\left({#1}\right)}
\newcommand{\pd}[2]{\frac{\partial{#1}}{\partial{#2}}}
\newcommand{\td}[2]{\frac{\rmd{#1}}{\rmd{#2}}}
\newcommand{\curl}{\nabla\times}
\renewcommand{\div}{\nabla\cdot}
\newcommand{\bB}{{\boldsymbol B}}
\newcommand{\vpl}{{\boldsymbol v}_{\rm pl}}

\newcommand{\skl}[1]{{\color{black}{#1}}}

\shorttitle{The game of life on a magnetar crust}
\shortauthors{Lander}

\begin{document}

\title{The game of life on a magnetar crust: from $\gamma$-ray flares to FRBs}

\author{S. K. Lander}
\email{samuel.lander@uea.ac.uk}
\affil{Physics, Faculty of Science, University of East Anglia, Norwich NR4 7TJ, U.K.}


\begin{abstract}
This paper presents a model to unify the diverse range of
magnetar activity, through the building and release of elastic
stress from the crust. A cellular automaton drives both
local and global yielding of the crust, leading to braiding of coronal
loops and energy release. The model behaves like a real magnetar in many ways: giant
flares and small bursts both occur, as well as
periods of quiescence whose typical duration is either
$\lesssim 1\,{\rm yr}$ or $\sim 10-30\,{\rm yr}$. The burst energy distribution broadly follows an
earthquake-like power law over the energy range
$10^{40}-10^{45}\,{\rm erg}$. The local nature of coronal loops allows
for the possibility of high-energy and fast radio bursts from the
same magnetar. Within this paradigm, magnetar observations can be used
to constrain the poorly-understood mechanical properties of the neutron-star crust.
\end{abstract}

\section{Introduction}

Magnetars, a class of restless neutron star characterised by energetic
outbursts, drive a wide range of astrophysical phenomena: from short
X-ray bursts of $\lesssim 10^{41}$ erg, through storms of bursts and
prolonged intermediate events, up to rare giant $\gamma$-ray
flares of $10^{44}-10^{46}$ erg, among the most violent events in the
Universe \citep{turolla,kas_bel}. Very recently, observations \citep{mereghetti,bochenek,chime_frb,li21,ridnaia}
have shown us that magnetars are also the
central engines for at least some fast radio bursts (FRBs);
that some short $\gamma$-ray bursts are the result of giant flares from extragalactic
magnetars \citep{burns}; and that certain ultralong-period radio emitters may be old
magnetars \citep{gleam,caleb,beniamini}.
In all its activity, the magnetar's solid crust plays a key
role: it stores an enormous amount of energy in the form
of elastic stress $\tau$ built up as the intense internal magnetic field $\bB$
evolves; seismic events then release some of this \skl{energy} into the 
corona, ultimately leading to the activity we observe \citep{ruderman91,TD95,perna_pons,L15,dehman}.

Compared with the detailed
quantitative simulations of crustal magnetic-field evolution
\citep{pons_vig,gour_symm}, our modelling of how the crust releases
elastic energy is rudimentary, and faces conceptual challenges \citep{T17}. Furthermore,
even the qualitative picture of magnetar activity is disjointed,
with e.g. short bursts and giant flares generally
treated as being of different physical origin, impeding any
  attempts to probe the underlying crustal physics.
Here, by contrast, we show how the full spectrum of magnetar activity can be
interpreted by taking a new approach: a single, physically-motivated model of the crust as
a cellular automaton that drives coronal activity and thus the observed bursting behaviour.

\section{Model}

\subsection{Crustal failure}

The \skl{outer crust of a magnetar, with a density $\rho<4\times
  10^{11}\,{\rm g\, cm}^{-3}$,} is
relatively weak and \skl{will be} partially molten for younger stars; although
  it may be the source of weak bursts \citep{younes22} we neglect its effect
  here, and concentrate on the more universal role of the inner crust,
an immensely strong crystalline structure that
resists and responds elastically to any imposed force, until it
reaches its elastic yield stress $\tau_{\rm el}$. 
When the crust eventually yields, the high pressure inhibits the
formation and propagation of voids through the crustal lattice, so
instead of a brittle fracture the crust is expected to flow
plastically \citep{jones03}, releasing elastic energy\footnote{Note, however, that in the model of
  \cite{T17} narrow crack-like plastic features develop.}.

\skl{To study crustal failure quantitatively, we will first need
  profiles of the mass and charge density $\rho,\rho_e$ and
  composition throughout the crust. These are found by solving the TOV
  stellar structure equations together with the SLy4 equation of state, as in
  \cite{LG19}. Fixing the mass at $1.4M_\odot$ gives us a model of
  radius $11.7\,{\rm km}$, whose inner crust is $550\,{\rm m}$ thick,
  which we adopt throughout this paper.}

\skl{To estimate the free energy reservoir of the crust, we evaluate the
formula for $\tau_{\rm el}$ from \citet{chug_hor} and
volume-integrate it over the inner crust, to find a maximum elastic energy}
$E_{\rm el}\sim\int\tau_{\rm el}\,\rmd V=4\pi\int\tau_{\rm el}(r)\,r^2\rmd r\approx 10^{47}\,{\rm erg}$.
 The energy of the
magnetic field threading the crust is comparable with $E_{\rm el}$; if
this is also tapped during crustal failure, the total
energy reservoir becomes $\sim 2\times 10^{47}\,{\rm erg}$.
This is a factor of $\sim 10$ greater than the most powerful known giant
flare \citep{palmer05}; crustal energy alone is therefore able, in
  principle, to explain all magnetar activity observed to date.

Crustal yielding is an essential part of the magnetar paradigm, as it
drives the transfer of energy, via the motion of embedded coronal
field footpoints, out to the corona, from where it is released
in the activity we observe \citep{lyu06}. 
A major challenge in neutron-star physics is how yielding occurs; microscopic molecular-dynamics simulations exhibit
collective local failure \citep{hor_kad}, but if every small group of crustal
ions were to yield as soon as $\tau=\tau_{\rm el}$, all resultant
bursts would be undetectably small. Equally, the crust's
stress distribution is likely to be highly anisotropic, with regions
where $\tau\approx\tau_{\rm el}$ and others with $\tau\ll\tau_{\rm el}$,
so it is energetically disfavoured for \skl{every} local failure to grow into a global one. 
Furthermore this would lead to a scenario, contradicted by observations, where a magnetar
would be unable to produce repeated small bursts. An additional
piece of physics must, therefore, set the characteristic lengthscale for crustal failure.


\begin{figure}
\begin{center}
  \begin{minipage}[c]{0.8\linewidth}
    \includegraphics[width=\linewidth]{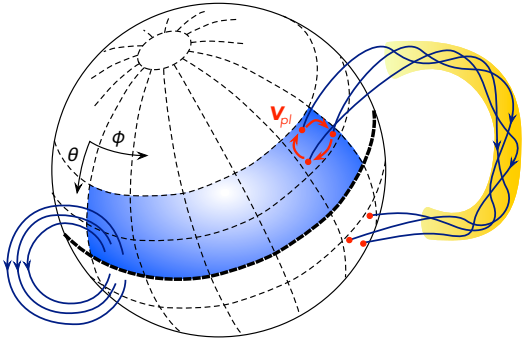}
  \end{minipage}
\caption{\label{coronacell}
The magnetar crust as an array of cells. The left-hand bunch of coronal field lines are untwisted,
embedded in a static cell in its elastic phase. The right-hand bunch
are braided by a plastic flow $\vpl$ circulating around a cell whose
elastic yield stress has been exceeded.}
\end{center}
\end{figure}

\subsection{The crust as an array of cells}
\label{cell_array}

First-principles macroscopic
simulations of crustal failure are currently out of reach, but there
are many clues to guide us to the origin of magnetar activity. Observationally,
numerous studies have shown how bursting activity appears to come from relatively
small patches of the crust \citep{palmer99,younes22}, often
at locations across the stellar surface
\citep{scholz_kas,younes20}, meaning that crustal failure must often be a local
phenomenon. Both high-energy bursts \citep{cheng96} and FRBs \citep{wadia_tim} exhibit a power-law
distribution of number vs energy, like that of earthquakes \citep{bak02}. From the
theory side, 3D numerical evolutions of 
the crustal magnetic field show the development of $\sim 1\,{\rm km}^2$ patches with strong
$B_\phi$ \citep{gour16,igoshev}; this significant change from the
initial $B$ induces high stress \citep{LG19}, and so
elastic failure is likely to occur \skl{within} such patches, but not necessarily spread beyond them. 
On the other hand, a crust-powered giant flare
requires a larger-scale failure, to explain the amount of energy released and how
much quieter the star becomes afterwards; the model must therefore
also allow for this possibility.

Guided by these considerations, we split the inner crust into an array of semi-autonomous
cells with fixed boundaries, each of surface area
$1\,{\rm km}^2$. To fix the cell depth, we note that
  since $\tau_{\rm el}$ increases by a factor of $\sim 1000$
  from the top to the base of the inner crust, we
  do not expect every local failure of a cell to propagate to the full
  $550\,{\rm m}$ depth of the inner crust. Instead we fix the cell depth at $200\,{\rm m}$, over which
  $\tau_{\rm el}$ varies by less than an order of magnitude, and
  assume that the crystalline structure in such a cell always fails
  collectively. \skl{As a result, we can ignore variations of
    the stress within a cell and assume it is given by a single spatially
    constant $\tau$ that evolves with time. This evolution is
    dependent on other parameters (see Eq. \ref{dtaudt}), so for consistency we therefore
    assume there is no spatial variation in \emph{any} physical quantity within a cell;
  we use the value of each from the base of the cell. Physical quantities do,
  however, vary from cell to cell across the crust.}

\skl{When a cell's stress exceeds the elastic yield value, $\tau>\tau_{\rm el}$, only a fraction of it
  is expected to be relieved in the ensuing plastic flow
\citep{LG19}; if we assume for definiteness a 10\% reduction
  from $\tau_{\rm el}$, the corresponding energy release is $3\times
  10^{40}\,{\rm erg}$ --}
similar to a fairly powerful short X-ray burst, and providing a
  sanity check of the cell model.

We regard a cell's stress as being sourced by the magnetic field $\bB$
alone, and define a scalar stress $\tau\equiv B^2/4\pi$. Neglecting the effect of
Ohmic decay -- reasonable for young magnetars \citep{pons_vig} -- the
evolution of $\bB$ in a crust stressed beyond $\tau_{\rm el}$ is
  dictated by an interplay of Hall drift and advection due to plastic flow $\vpl$:
\beq\label{B_evol}
\pd{\bB}{t}=-\curl\left[ \frac{c}{4\pi \rho_e}(\curl\bB)\times\bB \right]
                    + \curl(\vpl\times\bB).
\eeq
We model the crust as a Bingham
plastic: below $\tau_{\rm el}$ its response to stress is purely
elastic, with $\vpl={\bf 0}$ and only the Hall drift term present in
 \eqref{B_evol}, whilst above $\tau_{\rm el}$ the crust behaves as a
viscoplastic with flow velocity $v_{\rm pl}\propto(\tau-\tau_{\rm el})$.
Specifically, we use a scalar version of
the slow viscous-flow model of \citet{LG19},
\skl{produced by replacing spatial derivatives $\nabla\to 1/L$
(where $L$ is a characteristic lengthscale)}:
\beq\label{vpl}
v_{\rm pl}=\frac{L}{\nu}(\tau-\tau_{\rm el}),
\eeq
where $\nu$ is the viscosity of crustal matter in its plastic phase.
Now, from  \eqref{B_evol} we can derive an approximate scalar
  equation for $\tau$, by using the identity
$\partial B^2/\partial t=2\bB\cdot\partial\bB/\partial t$, and eliminating $B,v_{\rm pl}$
using  \eqref{vpl} and the relation $\tau=B^2/4\pi$:
\beq\label{dtaudt}
\td{\tau}{t}=\frac{c \tau^{3/2}}{\pi^{1/2}\rho_e L^2}
                    -\frac{2\tau(\tau-\tau_{\rm el})}{\nu},
\eeq
where we have swapped the signs on the two right-hand-side terms to
reflect the tendency of Hall drift to increase $\tau$ and
plastic flow to reduce it. 

Simulations \citep{LG19,GL21} show that stresses substantially higher
than $\tau_{\rm el}$ may form before plastic flow has a chance to relieve them;
to mimic this we model a cell's response as elastic, and allow $\tau$ to grow
  under the Hall effect, until it reaches a critical value 
  $\tau=1.1\tau_{\rm el}$. At this point failure occurs and in
  principle there will be an interplay between the Hall and plastic
  terms. To understand this, let us use  \eqref{dtaudt} to
  estimate characteristic timescales $\mathfrak{t}_{\rm
    Hall},\mathfrak{t}_{\rm pl}$ for the evolution of stress under
  Hall drift and plastic flow:
  \beq\label{timescales}
  \mathfrak{t}_{\rm Hall}=\frac{\pi^{1/2}\rho_e L^2}{c\tau^{1/2}}\approx 800\,{\rm yr}\ ,\ \
  \mathfrak{t}_{\rm pl}=\frac{\nu}{2(\tau-\tau_{\rm el})}\approx 9\,{\rm yr},
  \eeq
  using typical values:
  $L=200\,{\rm m},\tau=1.1\tau_{\rm el},\rho_e=1.4\times 10^{26}\,{\rm esu}\,{\rm cm}^{-3},\nu=10^{36}\,{\rm poise}$.
  The Hall effect is always active, but may reasonably be neglected
  during the plastic phase, since $\mathfrak{t}_{\rm pl}\ll\mathfrak{t}_{\rm Hall}$.
  During the elastic phase, only the Hall effect operates. Therefore, we model a single cell's evolution as
  alternating phases of growth of $\tau$ until the value
  $1.1\tau_{\rm el}$ is reached, followed by a reduction of $\tau$
  down to $\tau_{\rm el}$ under the action of plastic flow\footnote{More precisely, since $\rmd\tau/\rmd t\to 0$ as
  $\tau\to\tau_{\rm el}$, the plastic phase is ended
  at $\tau=1.0001\tau_{\rm el}$.}.

Time-integrating \eqref{dtaudt}, we
find closed-form expressions $\tau=\tau(t,t_{\rm swap})$ for both the Hall and
plastic phases, where $t_{\rm swap}$ is the time at which the current
phase began. The Hall phase takes $75$
yr, and the plastic phase $\sim 1-5$ yr -- somewhat shorter than
  the timescale estimates of  \eqref{timescales}, since here $\tau$ is
  only reduced by 10\%.

The crust's temperature $T$ has a minimum `ambient' value $T_{\rm amb}$ defined at the start of
each simulation. \skl{We wish to consider a range for $T_{\rm amb}$
  that encompasses all neutron stars that might plausibly display
  bursting activity, from young magnetars whose high surface
  $T$ appear to require a heat source in the crust
  \citep{belo_li}, to old sources ($>\!10^4\,{\rm yr}$) that have
  experienced no heating since birth. These considerations lead us to adopt, based on
  the cooling evolutions of \citet{hoglamand},
  the range\footnote{\skl{Note that the inner crust will be solid for
    all these models; it first begins to melt for $T>2\times
    10^9\,{\rm K}$.}} $T_{\rm amb,9}\equiv T_{\rm amb}/(10^9\,{\rm K})=0.05-0.5$.
Next,} $\vpl$ causes heating at rate $Q_{\rm pl}$, which
we model using an
approximation to equation (21) from \cite{lilevbel}:
\beq
\td{T}{t}=\frac{Q_{\rm pl}}{C_V}\sim \frac{B^2}{4\pi C_V}\frac{v_{\rm pl}}{L}
   =\frac{1}{\nu C_V}\tau(\tau-\tau_{\rm el}),
\eeq
integrating from $T=T_{\rm amb}$ at the start of the plastic phase,
and where $C_V$ is the specific heat capacity.
60\% of this heat is assumed to stay in the cell and the rest to
diffuse to its surroundings, to mimic the effect of thermal
conductivity. Once the plastic phase ends, $T\to T_{\rm amb}$ linearly
over a timescale $\approx C_V T/Q_\nu\approx 13$ yr (setting $T_9=1$
\skl{for the plastically-heated cell, and using values for $C_V$ and
  neutrino emissivity $Q_\nu$ from \cite{gnedin}, evaluated at the
  base of the cell as usual}).

Finally, we need an
expression for $\nu$. Since this is unknown from first
principles, numerical experimentation has been required to understand
the range of plausible values that produce an interplay between Hall
drift and plastic flow, and therefore allow for magnetar activity
\citep{L16}. Following \cite{LG19}, the density-dependence of $\nu$ is taken to be
the same as for $\tau_{\rm el}$. The $T$-dependence of viscous
fluids is often approximated by the Andrade equation \citep{andrade}, where
$\nu\propto{\mathrm e}^{1/T}$. Modifying this to avoid the divergent
behaviour (problematic for a solid) as $T\to 0$, and adjusting
constants to match previous work \citep{LG19,GL21,kojima}, we arrive at the
phenomenological relation
\beq\label{nu_prescr}
\nu(\rho,T)=5\times 10^{5} \tau_{\rm el}(\rho){\mathrm e}^{5/(1+T_9)}\ \textrm{poise},
\eeq
giving a possible range $10^{34}\lesssim\nu\lesssim 10^{36}$ poise for
a cell. 
Heating reduces $\nu$, affecting the crust's activity in two opposing
ways: on the one hand it increases $v_{\rm pl}$ and the rate of
coronal twisting; on the other hand it shortens the plastic phase and
reduces the chances of several plastic cells having time to join up
into a cluster, which in turn makes future deep failures and giant
flares less likely. Any other relation for $\nu$ of the same order of
magnitude, and reducing with $T$, would lead to broadly similar results.

\subsection{Cellular automaton}

Having described the physics of a single
cell, we need to understand how they interact.  We assume a cell's
behaviour only affects its four von Neumann neighbours
(i.e. those with whom it
shares an edge). It is well known that
complex physics can arise from simple cellular automata \citep{vonneumann,conway}.
At the same time, these rules need to be linked to underlying physics as rigorously as
possible to have any predictive power, and results cannot be artefacts
of a fine-tuned model, but should be robust and generic:
self-organised criticality (SOC) \citep{katz86,bak88}. There is
  evidence that both X-ray bursts and FRBs are driven by the same SOC
  process \citep{wei21}, motivating the present study.

Crustal magnetic field lines thread multiple cells, and are
dragged around locally by $\vpl$ in a cell. This exerts a
shearing force on its neighbours, but cannot cause them all to
fail -- otherwise every localised $\vpl$ could quickly
propagate across the entire crust. Instead, we encode this effect through a cell
rule: a cell in its elastic phase normally switches to its
plastic phase at $\tau=1.1\tau_{\rm el}$, but for every 
plastic neighbour, the cell's yield stress is lowered by
$0.025\tau_{\rm el}$; nearby plastic flow thus \emph{hastens}, rather
than triggers, a cell's failure. This is the key rule that leads to SOC-like
behaviour of the model. 
A contiguous cluster of plastic cells is regarded as a single physical entity, with
$\vpl$ circulating across the entire cluster with some
average velocity $\bar{v}_{\rm pl}$.

Shallow failures, down to the base of the cells, do not 
release enough energy to explain larger magnetar events, so the deeper
crust -- which will also be close to its $\tau_{\rm el}$ -- must also fail sometimes. Our
criterion for this to occur is that the cell itself, and at least three neighbours, must all be in a
plastic phase simultaneously. Such a `deep' failure below a cell
releases all stored elastic energy in that region, down to the
crust-core boundary. Thereafter we assume $\tau=0$ for that cell
(replenishing $\tau$ back to $\tau_{\rm el}$ through the entire
  inner crust would take substantially longer than the previous
  estimate \eqref{timescales} of $\mathfrak{t}_{\rm Hall}\approx 800\,{\rm yr}$); in this way, magnetar crustal dynamics are
analogous to forest-fire models \citep{dro_schw}.

\subsection{Corona}

The most readily induced kind of plastic flow
satisfies $\div\vpl=0$ and has no radial
component \citep{L16}. Restricting $\vpl$ to a cell, the only
permissible motion is a $\theta-\phi$ circulation of matter in loops
around the cell; the global crustal motion is inherently non-axisymmetric. The footpoints of external magnetic field
lines are embedded in the cell, so $\vpl$ causes a braiding of
these; see Fig \ref{coronacell}. $\bar{v}_{\rm pl}$
within a contiguous cluster of plastic cells plays two roles: it increases the average twist $\psi$ of the associated
coronal loop, $\rmd\psi/\rmd t=\bar{v}_{\rm pl}$, and is also taken to
represent the rate of transfer of elastic to coronal energy
$E_{\rm clus}$ for the cluster, so that:
\beq
E_{\rm clus}=\psi\brac{E_{\rm shallow}+E_{\rm deep}}.
\eeq
where $E_{\rm shallow},E_{\rm deep}$ are the sums of energy releases
from all shallow and deep failures, respectively.
The total coronal energy $E_{\rm corona}$ is then the sum of all $E_{\rm clus}$.

Magnetar bursts are linked to magnetic reconnection in the corona
\citep{lyu06}, a complex process that is not well understood; any
attempt to implement a detailed prescription risks introducing several
new poorly-constrained parameters, making the model
harder to constrain or falsify afterwards. Instead
we simply assume that when a plastic cluster ceases to exist, its
remaining associated $E_{\rm clus}$ is emitted as a single burst. In the special case
where at least one cluster's (average) twist reaches a peak value
$\psi>0.3$ rad, we impose a `high-twist' prescription
where all coronal braids reconnect, $\psi$ is reset to zero for each,
and the total $E_{\rm corona}$ is ejected at once in a `giant flare'.

\section{Numerical code}

To avoid conceptual issues where the two
footpoints of a coronal field line might both move in such a way that
$v_{\rm pl}$ does not cause any increase in $\psi$, we assume one
footpoint of every coronal field line is in the
`active' Northern hemisphere, and the partner footpoint in a `passive'
Southern hemisphere, releasing elastic energy but not driving the
motion. Covering the Northern hemisphere at a radius
$11.3\,{\rm km}$ (the outer boundary of the inner crust) requires
$\approx 800$ cells of $1\,{\rm km}^2$ surface area; we also
want a grid which is four times as long in $\phi$ as in $\theta$
(since $0\leq\theta<\pi/2,0\leq\phi<2\pi$). We therefore choose a fiducial resolution of
$14\times 56$, i.e. $784$ cells. For the top row of cells, identified with the pole, we set
$\tau=0$; those around the equator are
assumed to be mirrored (for the purposes of counting numbers of
plastic neighbours) with an unmodelled set of partner cells in the
Southern hemisphere. Periodic boundary conditions are imposed to
identify the $\phi=0$ and $\phi=2\pi$ edges of the grid.
The network of cells is evolved for $1000\,{\rm yr}$ with a C++ code which
tracks the formation, evolution and extinction of plastic clusters,
with a default timestep of $0.01$ yr. At the start of each simulation all cells are in the
Hall phase, with stresses in the range $0.9<\tau/\tau_{\rm el}<1.1$ randomly
assigned to each cell.
Thus, in this paper the time $t=0\,{\rm yr}$ represents a mature magnetar's crust at age $\sim 1000$ yr; the
newborn crust is unstressed, so no seismic activity will occur at
this stage. Differences in magnetic-field strength
and topology mean that some magnetars will reach this highly-stressed state earlier than
others, but are otherwise not likely to result in radically different
behaviour of the model.


\begin{figure}
\begin{center}
  \begin{minipage}[c]{0.9\linewidth}
\includegraphics[width=\linewidth]{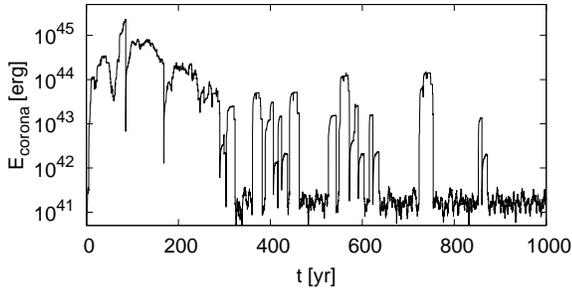}
\end{minipage}
\caption{\label{Ecorona}
First thousand years of evolution, for $T_{\rm amb,9}=0.1$. $E_{\rm
  corona}$ is the total twist energy of all coronal loops.}
\end{center}
\end{figure}

\begin{figure}
\begin{center}
  \begin{minipage}[c]{0.9\linewidth}
\includegraphics[width=\linewidth]{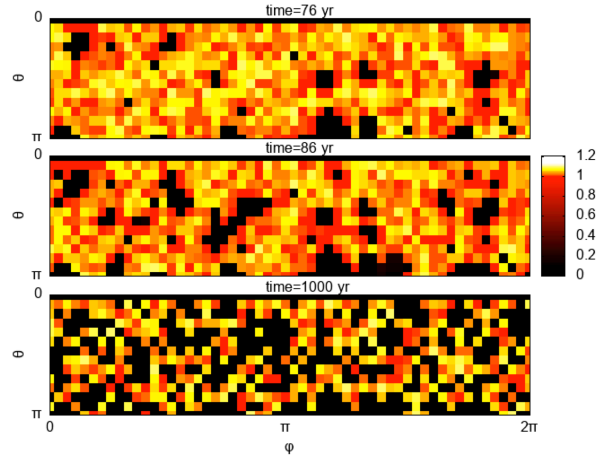}
\end{minipage}
\caption{\label{stresses}
Snapshots of $\tau/\tau_{\rm el}$ (colourscale) across the Northern
hemisphere, just before (top) and after (middle)
the first `giant flare', of $2\times 10^{45}\,{\rm erg}$, shown in the
previous figure. $\tau/\tau_{\rm el}$ at $1000$ yr (bottom) is also shown.}
\end{center}
\end{figure}

\section{Results}

Fig \ref{Ecorona} shows a representative example
light curve from $1000$ yr of evolution.
The first high-twist
`giant flare' event is seen at $t=85\,{\rm yr}$.
Fig \ref{stresses} shows the crust's stress pattern before and after this
event and demonstrates how every giant flare leaves
behind extended patches of unstressed crust, thus reducing the chances
of any future large-scale event \skl{occurring}. The fractal pattern remaining
after $1000$ yr is characteristic of these simulations, and is
seen at higher resolutions too.

\begin{figure}
\begin{center}
\begin{minipage}[c]{0.9\linewidth}
\includegraphics[width=\linewidth]{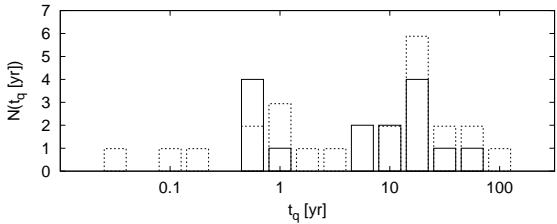}
\end{minipage}
\caption{\label{quiescence}
Number $N$ of quiescent periods of duration $t_q$ for two typical $1000$-yr
evolutions, at $T_{\rm amb,9}=0.05$ (solid bars) and $T_{\rm
  amb,9}=0.5$ (dashed bars).}
\end{center}
\end{figure}

$E_{\rm corona}$ can be high for long periods without any individual
loop developing high twist (although note that our simulations have no
term for twist decay); in such a state the magnetar could be
relatively quiet, but perhaps with a significantly
non-dipolar spindown rate (making estimates of the external field
unreliable) \citep{harding_CK,TLK}. Long periods of
`quiescence' (defined here as $E_{\rm corona}<10^{42}$ erg) punctuated by
occasional intermediate events are seen, especially at later
times. Fig \ref{quiescence} shows durations of these quiescent periods, for two representative runs
with different $T_{\rm amb}$. The distribution is roughly bimodal: our model magnetars go quiet
for either a few months, or $\sim 10-30$ yr. 
There is no correlation between the size of smaller events
and the waiting time until the next one; though in the aftermath of a
giant flare the model stars are -- like real magnetars -- often quieter.

The focus of this paper is on understanding the crustal dynamics
  and transfer of energy to the corona, a widely accepted idea for
  high-energy magnetar bursts that is now also a leading model to explain (at
  least some) FRBs. Radio emission, however, is likely to require
  substantially lower levels of coronal twist than high-energy
  emission \citep{wadia_tim}, making it hard to reconcile the standard
  globally-twisted magnetar corona model (e.g. \cite{TLK}) with recent
  observations of contemporaneous FRBs and X-ray bursts from the
  magnetar SGR 1935+2154 \citep{mereghetti,bochenek,chime_frb}. By
  contrast, this paper's model, where crustal failure leads
  to a network of more localised coronal braids with varying levels of
  twist, naturally allows for this.

  Whilst we cannot determine whether a given
  event from our evolutions will ultimately be seen as an
  X-ray burst or an FRB, since we do not study emission physics, we
  can still infer whether conditions in the corona are propitious for
  generation of a particular kind of radiation. From
  Fig. \ref{Ecorona} we see that for the first $300\,{\rm yr}$ the
  corona always contains a lot of twist energy, making FRB emission
  highly disfavoured. Confirming observational results \citep{lin20},
  we thus expect a classically active young magnetar to produce far more
  X-ray bursts than FRBs. Later on in our evolutions, energetic bursts of
  shorter duration occur, in between increasingly long periods of
  quiescence, and we thus anticipate -- following \cite{wadia_tim} --
  that FRB emission will become more likely.

\begin{figure}
\begin{center}
\begin{minipage}[c]{1.0\linewidth}
\includegraphics[width=\linewidth]{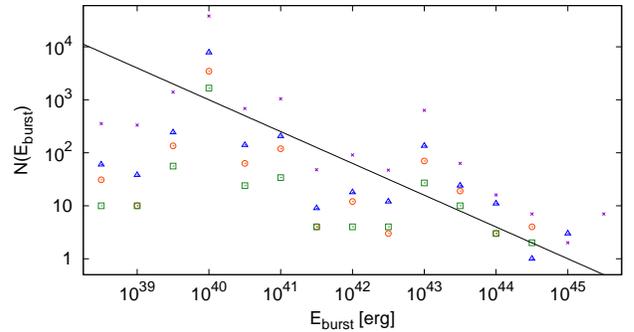}
\end{minipage}
\caption{\label{burst_4res}
Number of bursts $N(E_{\rm burst})$ with energy $E_{\rm burst}$
  over a $1000$-yr simulation, in
$\log(E_{\rm burst})=0.5$ bins, compared with a power-law relation of
$\Gamma=1.6$ (the line shown). Fixing a temperature $T_{\rm
  amb,9}=0.5$, we explore the effect of cell size. Results for resolutions $14\times 56$
  (fiducial), $10\times 40$, $20\times 80$, $44\times 176$ are indicated with the
  $\circ,\Box,\vartriangle,\times$ points, respectively.}
\end{center}
\end{figure}

Fig \ref{burst_4res} plots the burst energy
  distribution from $1000\,{\rm yr}$ of simulation, showing that burst
  numbers $N(E_{\rm burst})$ in the energy range $E_{\rm burst}=10^{40}-10^{45}$ erg broadly follow a
Gutenberg-Richter \citep{GR_law} power law
$\rmd N/\rmd E\propto E^{-\Gamma}$ \skl{independent of cell size. The
  total number of events does however increase with resolution,
  because there are more cells available to undergo elastic
  failure. Note that the evolution with $44\times 176$ cells, ten
  times the fiducial value, is shown as an extreme case; the other
  three resolutions are likely to be more realistic (recall Section \ref{cell_array}).}
The infrequency of $E_{\rm burst}<10^{40}$ erg events is an
artefact of our model, which considers the inner crust only;
lower-energy bursts are likely to involve the outer crust.
Our model predicts that no magnetar will produce more than $\sim 10$
\skl{giant flares} over its first $1000\,{\rm yr}$ of maturity and -- given the results of Figs
\ref{Ecorona} and \ref{stresses} -- none at all thereafter.

Cellular automaton models, and many different kinds of
astrophysical source, generically exhibit power-law energy
distributions with $\Gamma\approx 1.5-2$ \citep{aschw16}; in the
fractal-diffusive model for cellular automata the key variable is
the spatial dimension of the cell dynamics, with $\Gamma=1.5$
predicted for 3D models \citep{aschw12}. This is very close to the typical value
$\Gamma=1.6$ for magnetar X-ray bursts
\citep{cheng96,gogus99,gogus00,gavriil}, which is plotted in Fig
\ref{burst_4res} for comparison.

In corona-type cellular automata models, the cell rules encode an
immediate diffusive redistribution of
energy from one cell to its neighbours \citep{isliker} and lead to a tight
$N-E$ correlation \citep{lu_ham,danila}. Here the cell rules have the less
direct effect of making energy release from a plastic cell's
neighbours \emph{more likely} rather than guaranteed, and as a result the
burst statistics show more deviation from a simple power law. There
is also an overrepresentation of $10^{40}$ and $10^{43}$ erg events
(corresponding to a single cell's shallow or deep failure,
respectively). Within the paradigm presented here, therefore,
the degree of scatter of bursts from a power-law relation may
encode valuable information about the nature of magnetar crustal failure.

\section{Outlook}

The model presented here was constructed based on
theoretical considerations, and aims to be a faithful minimal representation of the
salient physics of magnetar activity. Where possible, the model is quantitative: the crustal structure is
calculated with a realistic equation of state, the
elastic stress from a fit to molecular dynamics simulations, and the
characteristic lengthscale for the cells taken from 3D magnetoelastic
simulations. The details of how the crust fails are, however, unknown,
which makes the use of somewhat ad-hoc prescriptions
inevitable. Nonetheless, we have performed extensive checks to confirm
that the model's \skl{key} results are robust to
variation in cell size (see Fig. \ref{burst_4res}), the expression for
$\nu$, and the critical value at which the transition
between elastic and plastic regimes occurs. In particular, we have not attempted to `tune' any
input quantities in order to better mimic the behaviour of any particular magnetar. With this
concrete model for crustal failure, however, we can now directly use
information from magnetar bursts to constrain the
model and the star's physics. For example, because even the largest
events in our current model are relatively localised and weaker than
the brightest known giant flare, this indicates the presence of an
additional mechanism driving the propagation of crustal failure,
e.g. a thermoplastic instability \citep{bel_lev}; this may be
manifested observationally as deviations from the power law for weaker
bursts. The detection of a giant flare with energy $\gtrsim
  10^{47}\,{\rm erg}$ would point to the
involvement of core magnetic field evolution.

With a temperature-dependent plastic viscosity, hotter
crusts produce frequent bursts, but slightly cooler ones are needed
to allow time for a large plastic cluster to form and
potentially power a giant flare. If the crust is too cold, however,
the sluggish plastic flow is more likely to lead to long-lived
multipolar coronal fields or non-dipole spindown than outbursts. It
will also be easier to observe FRBs from cooler crusts, since the more
sparsely distributed coronal loops will not significantly inhibit radio emission
\citep{wadia_tim,suvorov}.

Very young and very old magnetars are
not expected to produce giant flares: the former because large regions
of high stress have not yet developed; the latter because numerous
previous events have produced a fractal low-stress region that
inhibits further large-scale failure.

\skl{Magnetar activity is often linked directly to the evolution of
  the star's toroidal $\bB$, and although results of such simulations
  also inform our choice of cell geometry, our focus is instead on
  the distribution of evolving elastic stress in the crust. This
  evolution is driven by the changing $\bB$, but is insensible to its quantitative features. It is not obvious whether
  activity driven by an intense \emph{poloidal} $\bB$ would be
  noticeably different within our paradigm; perhaps it would lead to a
  different natural cell geometry. Any other source of stress
  could also, in principle, drive seismic activity, whether or not it then leads to characteristic
  magnetar behaviour. The most obvious example would be stresses
  developing through spindown; in this case both the cell geometry and
  the evolution equations would need to be revised.}

The crust is the \'eminence grise of magnetars: it powers their
activity, but in a way that is difficult to discern from
observations, which essentially `see' only the corona. The goal of
this work is to provide a framework to compare observations directly
with theory, and so to probe the crust's
poorly-understood mechanical properties.

\acknowledgements

It is a pleasure to thank Ersin G\"o\u{g}\"u\c{s}, George Younes and Zorawar
Wadiasingh for many stimulating discussions related to this work.


\bibliographystyle{aasjournal}

\end{document}